\documentclass[12pt]{article}
\usepackage{amssymb}
\usepackage{amsmath}
\newtheorem{ass}{Assumption}
\newtheorem{theorem}[ass]{Theorem}
\newtheorem{lemma}[ass]{Lemma}
\newtheorem{definition}[ass]{Definition}
\newtheorem{corollary}[ass]{Corollary}

\begin{document}

\title{A framework for perturbations and stability
of differentially rotating stars}

\author{Horst R. Beyer \\
 Max Planck Institute for Gravitational Physics, \\
 Albert Einstein Institute, \\
 D-14476 Golm, Germany}

\date{\today}                                     

\maketitle

\begin{abstract}
The paper provides a new framework for 
the description of linearized adiabatic lagrangian perturbations 
and stability of differentially rotating newtonian stars. In doing
so it overcomes problems in a previous framework by 
Dyson and Schutz and provides the basis of a rigorous analysis
of the stability of such stars. For this the 
governing equation of the oscillations is written as a first 
order system in time. From that system the generator of time evolution is 
read off and a Hilbert space is given where it 
generates a strongly continuous {\it group}. As a consequence the
governing equation has a well-posed initial value problem.
The spectrum of the generator relevant for stability considerations
is shown to be equal to the 
spectrum of an operator polynomial whose coefficients 
can be read off from the governing equation. 
Finally, we give for the first time sufficient criteria
for stability in the form of inequalities for
the coefficients of the polynomial. 
These show that a negative canonical energy of the star does not 
necessarily indicate instability.
It is still unclear whether these criteria are strong enough to
prove stability for realistic stars.  
\end{abstract}


\section{Introduction}
The study of oscillations of stars is an important and exciting field of 
current astrophysics. For instance through period-luminosity and period-radius
relationships variable stars provide important `yardsticks' 
for distance measurements in the universe. Their observation yield
important information about the interior of stars,
like the equation of state of the matter, which is otherwise hard to obtain.
Further, Neutron star pulsations may be a source of
gravitational radiation detectable for 
experiments like LIGO, VIRGO and GEO 600 in the near future.
\newline
\linebreak 
On the other hand, it is probably fair to say that there has not 
been very much work on the mathematical foundations of the theory
of stellar oscillations.
In the non relativistic limit there is a well-known framework for the 
description of oscillations of nonrotating stars
\cite{cox}, \cite{unno} (see \cite{beyer1} for a rigorous  version).
It is important to note that even on that level it turns out that the 
governing operator of the spheroidal oscillations belongs to a
class of operators which were apparently (apart from special 
cases considered in \cite{eisenfeld}) previously unconsidered in operator 
theory \cite{beyer3}. Somewhat surprisingly it also turned out that 
differently to radial oscillations these operators don't 
have a compact resolvent. Hence the well developed perturbation theory 
for such operators cannot be applied and a corresponding  
theory for the new type of operators still has to be developed. 
For rotating stars there is little known about the relevant
operators apart from abstract properties (like the symmetry, semiboundedness
and continuity of associated operators \cite{lyndenbellostriker,hunter}), 
indication of a continuous part in the spectrum 
\cite{kojima,beyerkokkotas} and instabilities caused by so called `r-modes' 
(also called `quasi-toroidal modes') \cite{anderssonkokkotasschutz,
lindblomowenmorsink}.
To my knowledge there is no consideration of these 
operators in sufficient detail. Indeed, a large part of
the present paper considers the more 
modest first step of such an investigation, namely to {\it identify} and determine 
which operators {\it should be} considered in that case.   
\newline
\linebreak
The governing equation for linearized
adiabatic oscillations of a stationary 
differentially rotating perfect-fuid 
star in an inertial frame $(t,x)$ is \cite{lyndenbellostriker}
\begin{equation} \label{basiceq}
\frac{\partial^2 \xi }{\partial t^2} +
B^{\, \prime} \, \frac{\partial \xi }{\partial t} +
C^{\, \prime} \, \xi = 0 \, \, , 
\end{equation}
where 
\begin{equation}
B^{\, \prime} \, \frac{\partial \xi }{\partial t} :=  
2 \, (v \cdot \nabla) \, \frac{\partial \xi }{\partial t} \, \, ,
\end{equation}
\begin{eqnarray} \label{Aprime}
C^{\, \prime} \, \xi := 
&-& \frac{1}{\rho} \, 
\nabla \left( p \, \Gamma_1 \nabla \cdot \xi \right) 
+ (v \cdot \nabla)^2 \xi \nonumber 
+ \frac{1}{\rho} \left[ 
- \nabla \left( (\nabla p) \cdot \xi \right)
+ (\nabla \cdot \xi) \, \nabla p \right] \nonumber \\
&-& \nabla \Phi_{\xi} + 
\sum_{j,k=1}^{3} \left( 
\frac{1}{\rho} \, \frac{\partial^2 p}{\partial x_j \partial x_k}  
+ \frac{\partial^2 \psi }{\partial x_j \partial x_k} \right) 
\xi_k \, e_j   \, \, , 
\end{eqnarray}
\begin{equation}
\Phi_{\xi}(t,x) :=    
- G \int_{\Omega} \frac{[\nabla \cdot (\rho \xi)](t,y)}{|x-y|} \, 
d^{\,3}y \, \, , \, \, 
\psi(x) :=    
- G \int_{\Omega} \frac{\rho(y)}{|x-y|} \, d^{\,3}y \, \, , \, \,
x \in \Omega \, \, ,
\end{equation}
$\xi$ is the Lagrangian displacement 
vector field, for $j \in \{1,2,3\}$ the symbol $e_j$
denotes the canonical unit vector in the direction
of $x_j$, $\Omega$ is the (bounded open) volume of the star, 
and $v,p,\rho, \Gamma_1$ are 
the velocity field, pressure, density and the adiabatic index 
functions of the background star satisfying 
the equations of momentum and mass conservation
\begin{equation}
(v \cdot \nabla) \, v = - 
\left( \frac{1}{\rho} \, \nabla p + \nabla \psi \right) 
\, \, , \, \, 
\nabla \cdot (\rho v) = 0 \, \, .
\end{equation}
and an equation of state.
In addition to (\ref{basiceq}) the variation $\delta p$ of the pressure
has to vanish at the boundary $\partial\, \Omega$ of the star, i.e.,
\begin{equation} \label{boundcond}
\lim_{y \rightarrow x} \left( \,
p \, \Gamma_1 \nabla \cdot \xi \right)(y) = 0 
\end{equation} 
for all $x \in \partial \, \Omega$.
\newline

The remarkable paper \cite{dysonschutz} of Dyson and Schutz
provides a framework for deciding the stability of the solutions 
of (\ref{basiceq}). In the following this paper is referred 
to as DS. Compared to previous 
frameworks given in \cite{lyndenbellostriker},
\cite{hunter} the main step 
forward in that paper is the fact
that it relates the stability of the system  
directly to the growth properties of the perturbations in time
as is usual for nonrotating stars (see, e.g., 
\cite{beyer1}, \cite{beyer2}, \cite{beyerschmidt}, 
\cite{cox}, \cite{ledouxwalraven}, \cite{unno}).
Moreover the paper shows that these growth properties 
are governed by spectral properties of the generator 
of time evolution. This greatly simplifies the stability
discussion. Unfortunately, the approach
still has some drawbacks, and in the present  
paper will be given a varied framework which 
overcomes those problems. Moreover here are given for the 
first time sufficient criteria for stability in the form
of inequalities which have to be satisfied by the 
coefficients of an operator polynomial. These criteria
show that a negative canonical energy does not 
necessarily indicate an instability of the star. It  
is still unclear whether these criteria are strong enough to
prove stability for realistic stars.
\newline
\linebreak
A rough 
discussion of the approach of Dyson and Schutz is now given. 
The paper considers axisymmetric solutions of the form
\begin{equation} 
\xi(t,x) = exp(im\varphi) \, \xi_m(t,r,\theta) \, \, , 
\end{equation}
where $r,\theta,\varphi$ are spherical coordinates 
and $m \in {\Bbb{Z}}$. Inserting this ansatz into 
(\ref{basiceq}) leads to an equation of the same 
structure with induced operators  $B_m^{\, \prime}$
 $C_m^{\, \prime}$. The index $m$ is supressed in the following
discussion.
A Hilbert space $H^{\, \prime}$ (here $X$) for the data is chosen 
such that, both, $B^{\, \prime}$ becomes {\it continuous
and antisymmetric} and 
$C^{\, \prime}$ becomes symmetric. In the nonrotating limit
$H^{\, \prime}$ goes over into the usual Hilbert space used 
in the stability discussion for spherically symmetric 
stars. A physically 
reasonable condition on the background model 
is given which leads to a lower bounded 
$C^{\, \prime}$. Assuming that condition 
$C^{\, \prime}$ is substituted by its so called
Friedrichs extension. This is an abstractly defined 
self-adjoint extension which exists for {\it every} densely defined
linear symmetric and semibounded operator in Hilbert space
(see e.g. \cite{reedsimon} Vol. II). \footnote{ The importance of 
choosing a {\it self-adjoint} extension of $C^{\, \prime}$ 
can be seen in the limit of no rotation. In \cite{beyer1}, 
\cite{beyer2} it is shown that for polytropic stars with 
a polytropic index $n < 1$ there is an infinite number of
different self-adjoint extensions which all lead to a
well-posed initial value problem for the wave equation.}    
In the standard way the resulting wave equation is written 
as a first order system in time for $\xi(r,\theta,t)$ and 
$(\partial \xi / \partial t)(r,\theta,t)$. The 
initial value problem of the system is studied. 
The Hilbert space of the data is chosen 
as $H^{\, \prime \, 2}$ with the induced
`euclidean' scalar product. However it is noticed that 
this is physically not meaningful, because the scalar product has no 
physical interpretation and is 
not even dimensionally correct. From the first order system 
the linear operator $T$ generating time evolution
is read of and it is shown that its spectrum 
is equal to the spectrum of a quadratic operator 
polynomial generated by $B^{\, \prime}$ and $C^{\, \prime}$
\cite{markus,rodman}.
Moreover the resolvent of $T$ can be given in terms of the 
inverses of the operator polynomial. That information 
along with estimates on the resolvent of the operator 
polynomial are used to give an estimate on the spectrum of 
$T$. In general that estimate is not strong enough 
to decide the question 
of stability of the system. From these estimates it is 
further shown that 
there is a solution of the initial value problem for 
the system corresponding to 
elements of the domain of $T$. The uniqueness of the solution 
is not shown. The authors remark that
they could not show that $T$ is the 
generator of a strongly continuous semigroup and as a 
consequence the results of standard semigroup theory
could {\it not} be used.  
Finally, the completeness of normal modes of the
system is discussed. \newline

From the description the reader might have noticed 
that in the derivation of these results 
only abstract properties like `continuity', `symmetry', 
`semiboundedness' and `self-adjointness' of $B^{\, \prime}$ 
and $C^{\, \prime}$ play a role. This is indeed true and is 
the reason why that approach is called here a `framework'. 
The same also applies to the approach here. 
As a consequence these frameworks
can be used to describe a lot more physical systems than stellar
oscillations. The main ingredient for such an application
is a system of wave equations which is second order
in time (with or without first order time derivatives) 
and which is not explicitly time dependent. 
The `coefficients' in that system can be (not necessarily local) 
linear operators with certain abstract properties. For this 
reason we abandon in Section 2 any reference 
to rotating stars 
and just consider {\it abstract} wave equations of type 
(\ref{basiceq}). 
Having this 
in mind might also provide a better understanding of
some of the statements below.
\newline

After this digression the discussion
of DS is continued. The main problem
of the approach comes from the chosen 
Hilbert space along with a scalar product which 
is not related to any physical quantity and not 
dimensionally correct. Of course the latter could be 
remedied by first introducing a dimensionless time coordinate.
But experience tells that this should not be essential 
at such an early stage. Also it 
is known that the use of a suitable Hilbert space
decides whether semigroup theory can be applied or not. 
So it is very likely that the use of $H^{\, \prime \, 2}$
is responsible for the fact that semigroup theory could not
be applied. Indeed a different choice of the 
Hilbert space will turn out to be the key to the results
of this paper. 
\newline

Another point which was not addressed in 
DS
is the fact 
that in addition to (\ref{basiceq}) the boundary 
condition (\ref{boundcond})
has to be satisfied that the Lagrangian variation 
$\delta p$ of the 
pressure vanishes at the surface of the star. 
\cite{ledouxwalraven,cox} Indeed it has been shown in 
\cite{beyer1,beyer2} for the limit of no rotation that 
for a polytropic equation of state with polytropic index $n < 1$
there is a infinite number of different self-adjoint extensions of 
$C^{\, \prime}$, which all lead to different initial value formulations
for the wave equation. Moreover it has been shown 
that the condition of a vanishing $\delta p$ at the surface of the star 
picks exactly one of these 
self-adjoint extensions. Of course the choice of the 
Friedrichs extension of $C^{\, \prime}$, is equivalent to
posing a boundary condition. But because of the abstractness 
of this extension it is not obvious
and has to be investigated whether 
it is compatible with (\ref{boundcond}). To my knowledge this 
has been shown only for the case of {\it radial} oscillations of 
spherically symmetric stars in \cite{beyer1}. This point 
will not be pursued any further in this paper.
\newline

The approach in this paper is similar to that of Dyson 
and Schutz. The point of departure is in  
the choice of the Hilbert space for the 
initial data of the first 
order system. Here a space $Y$ is chosen 
, which is in general a proper subspace 
of  $H^{\, \prime \, 2}$. Moreover a different
and dimensionally correct scalar product is chosen. 
The square of the induced norm of the 
initial data is a positive definite part
of the corresponding canonical energy of the 
system.\cite{friedmanschutza,friedmanschutzb} 
For this $C^{\, \prime}$ is split into sum of
a strictly positive self-adjoint operator $A$ and
a `rest' $C$. Of course such a decomposition is not unique
but it can be shown (see Lemmas 14 and 15 in Section 2) 
that trivial rescalings all lead to the same set $Y$ along with 
equivalent norms on $Y$. In particular such changes lead 
to theories which are related by a
similarity transformation and hence the outcome of the 
stability discussion is not affected.
In general the canonical energy {\it cannot} be used 
as a norm for $Y$ because it is not always positive definite. 
In situations where it is 
$C$ can be chosen as zero.  
In the limit of no rotation where $B^{\, \prime} = 0$ and the operator 
$C^{\, \prime}$ is semibounded the approach here
reduces to the approach in \cite{reedsimon} Vol. II 
(see the proposition at the beginning of chapter X.13) for 
classical wave equations.
\newline

A major consequence of the change is that it allows the 
use of semigroup
theory which is a standard and well developed 
tool in particular in the theory of partial 
differential equations (see e.g. \cite{davies},\cite{engel},\cite{hillephillips},
\cite{goldstein},\cite{pazy}, \cite{renardyrogers}
and the cited references therein). This simplifies the reasoning
a lot, because it can be and will be 
built on those results. In particular here the operator
$G_{+}$ which corresponds to $T$ in DS
generates a 
strongly continuous {\it group} of bounded transformations 
and hence the well-posedness of the initial value problem 
for the first order system 
follow from abstract semigroup theory.  
At the same time a considerable generalization 
is achieved. The operator $B^{\, \prime}$ (here denoted by 
$iB$) can be unbounded
and not antisymmetric. Moreover
$C^{\, \prime}$ has not to be assumed  
symmetric. The restrictions 
imposed on these operators are the following. 
The operator $C^{\, \prime}$ has to be of the form 
$A + C$ where $A$ is some densely defined 
and strictly positive self-adjoint operator
in $X$ and $C$ is a relatively bounded perturbation of the positive 
square root $A^{1/2}$ of $A$.
In addition $B^{\, \prime}$ has to be a relatively bounded 
perturbation of $A^{1/2}$ 
with relative bound smaller than $1$. Finally, 
$B^{\, \prime}$ has to be antisymmetric or continuous, but not 
necessarily both. All these conditions
are trivially satisfied for the case of axisymmetric solutions
of (\ref{basiceq}) considered  
by Dyson and Schutz. 
Whether this generalization is sufficient to 
provide a framework for (\ref{basiceq}) and not only for
its axisymmetric form is not 
yet clear. For this it seems necessary that  
$C^{\, \prime}$ given by (\ref{Aprime}) 
is semibounded and this is still open.
The reason for considering also
more general situations with
nonantisymmetric $B^{\, \prime}$ and 
nonsymmetric $C^{\, \prime}$ is that
the framework here will also be used in a 
future paper 
in the stability discussion of the Teukolsky 
equations on a Kerr background where this is the case.
\cite{teukolsky}     
A further important advantage
of the approach here is that it can be shown (see Theorem 3)
that the
dominating part
of $G_{+}$ (but in general not $G_{+}$ itself) 
is {\it self-adjoint}. 
Using perturbation theory
this gives important information on the spectrum of $G_{+}$
and is also the basis of the proof that $G_{+}$ is the 
generator of a strongly continuous group of bounded 
transformations. (See Theorem 7) 
Further it is the basis for another result 
(see Corollary 12) having no 
counterpart in DS 
namely the conservation of the `canonical energy' $E$.
On the other hand it turns out 
that the spectrum of  $G_{+}$ is the 
{\it same} as of $T$. In particular that spectrum 
is given by the spectrum of the {\it same} 
operator polynomial $C^{\, \prime} - \lambda B^{\, \prime} + \lambda^2,
\lambda \in {\mathbb{C}}$ (See Theorem) 
\newline

A plausible definition for the stability of a rotating star
is the following. The system is stable 
if and only if the semigroup 
$T_{+}(t), t \in [0,\infty)$ generated by $G_{+}$ is bounded.
Note that this definition is invariant to
similarity transformations and hence not so sensitive to changes  
of Hilbert space like one only invariant 
under unitary transformations.\footnote{Such a definition would be given 
for instance by the demand that the semigroup should be {\it contractive}, i.e.,
that the norms of the semigroup elements are $\leqslant 1$.} 
From semigroup theory one has then the  
following.
\begin{enumerate}
\item 
The system is {\it unstable} if $G_{+}$ has a spectral value 
with real part smaller than zero.
\item 
For a {\it stable} system the corresponding spectrum of 
$G_{+}$ is contained in the closed right half plane of the complex
plane.
\item
From only the fact that the spectrum of $G_{+}$ is part 
of the closed right half plane of the complex plane,
it {\it cannot} be concluded that the system is stable.
\footnote{ For a counterexample compare for instance 
the note after the proof of Corollary 9.}
\item
The system is {\it stable} 
if the {\it real part} of all `expectation values'  
\begin{equation} \label{inequality}
(\xi | G_{+} \xi) 
\end{equation}
is positive ($\geqslant 0$) for all elements $\xi$ from the 
domain (or a core)
of $G_{+}$.\footnote{Then $G_{+}$ generates a {\it
contraction} semigroup.}
\end{enumerate}
Point 1 gives a sufficient but not necessary 
condition for instability. Note that this condition is invariant 
under similarity transformations. Moreover because of 
Theorem 13 it is equivalent to the condition that 
there is complex number $\lambda$ with real part smaller than zero such 
that 
\begin{equation} \label{oppencil}
C^{\, \prime} - \lambda B^{\, \prime} + \lambda^2
\end{equation}
is not bijective. This reduces in the nonrotating case 
to the condition that $C^{\, \prime}$ is strictly negative, which is 
a well known sufficient condition for instability.\cite{beyer1}
An important final observation is that from the existence of such
a $\lambda$ follows the existence of an element 
$\xi$ from the Hilbert space such that the corresponding
function of norms $|T_{+}(t) \xi|, t \in {\Bbb{R}}$ 
grows exponentially for large times.\footnote{This is easily seen 
for instance by using Theorem 4.1 in chapter 4 of \cite{pazy}. Here it is 
important to remember that in general the spectrum
of $G_{+}$ does not only consist
of `eigenvalues' (for which this statement is 
of course trivially satisfied) but also values $\mu \in {\Bbb{C}}$
for which the map $G_{+} - \mu$ is just not onto. Such values are often from 
a continuous part of the spectrum.} Hence the existence of 
such a $\lambda$ leads to a much stronger kind of instability. 
\newline 
\linebreak
Point 4 gives a sufficient but not necessary 
condition for stability. It is appealing because 
it is of the form of an   
inequality, 
which is more easily accessible than the spectrum of $G_{+}$.
On the other hand it is strong and
{\it not} invariant to similarity transformations.
It turns out to be equivalent to   
$C^{\, \prime}$ being strictly positive, i.e., that 
the spectrum of this operator consists only of positive real numbers 
different from zero. Note that this reduces 
to a known sufficient condition for stability in the nonrotating case. 
But for such stars it can be satisfied only for
radial oscillations (for instance this is the case  
for constant $\Gamma_1 > 4/3$), but not for nonradial
oscillations.\footnote{This is obvious since
the spectrum of the trivial toroidal oscillations is $\{0\}$.
But in \cite{beyer2} it has also been shown that $0$ is in the 
spectrum of spheroidal oscillations.}
\cite{beyer1,beyer2,beyerschmidt}  
Note that in the limit of no rotation the trivial toroidal
oscillations give rise to solutions of (\ref{basiceq},\ref{boundcond}) 
whose norm increases linear in time
for large times. Hence applying the stability definition 
above to that limit would lead to an `unstable star'. Of course,
these oscillations can be excluded in that case 
just by considering a reduced operator. 
\newline
\linebreak
The following two stability criteria are new. 
They will be proven 
in Theorem 17. 
\begin{enumerate}
\item
If $B^{\, \prime}$ and $C^{\, \prime}$ are such that
\begin{equation} \label{stab1}
<\xi|C^{\, \prime}\xi> - \frac{1}{4} <\xi|B^{\, \prime}\xi>^2 \, \, \geqslant 0
\end{equation}
for all $\xi$ from the domain of $C^{\, \prime}$ such that $\|\xi\|=1$
then the spectrum of $G_{+}$ is purely imaginary.
\item
If the operator
\begin{equation}
C^{\, \prime} - \frac{ib}{2} B^{\, \prime} -\frac{b^2}{4}
\end{equation} 
is positive for some $b \in {\mathbb{R}}$
then the spectrum of $G_{+}$ is purely imaginary and there 
are $K\geqslant 0$ and 
$t_0 \geqslant 0$ such that 
\begin{equation} 
|T_{+}(t)| \leqslant Kt  
\end{equation}
for all $t \geqslant t_0$. 
\end{enumerate}
Note for the first point that 
$-(1/4)<\xi|B^{\, \prime}\xi>^2$ is {\it positive}, because
of the antisymmetry of $B^{\, \prime}$. Also note in this 
connection that in DS it has been 
shown that $C^{\, \prime} - (1/4)B^{\, \prime \, 2}$ is bounded
from below {\it uniformly in m}. Unfortunately, in general 
this does not imply (\ref{stab1}). 
\newline

It is still unclear whether these criteria are strong enough to 
prove stability for realistic stars. On the other hand
the second criterium has been sucessfully applied in the stability 
discussion of the Kerr metric where the master equation governing 
perturbations is of the form (\ref{basiceq}), too.\cite{beyer4}

\section{The framework}

This section developes the initial value formulation for
abstract differential equations of the form (\ref{basiceq}). It 
is self-contained and necessarily very technical. The reader 
who is not 
interested in the excessive mathematical details given here is referred
to the introduction. The used nomenclature can be found in standard 
textbooks on Functional analysis.\cite{reedsimon} Vol. I,
\cite{riesznagy,yosida} 
\newline
\linebreak
Before going into the mathematical details it is explained about
the meaning of the individual results of this section.
The section is based on the assumptions 
General Assumption 1 and General 
Assumption 4 on three operators $A$, $B$ and $C$. 
A different form of General Assumption 1 which is 
more convenient for applications can be given in the 
obvious way using Lemma 18. Definition 2 gives the Hilbert space 
$Y$ which is used here instead of the Hilbert space  
in DS. A rigorous form (\ref{waveeq}) 
of (\ref{basiceq}) along with the existence and uniqueness of the solution
corresponding to initial values is given in Theorem 11. 
Corollary 12 gives the corresponding `energy' along with
an identity for its time derivative. 
The analogue $G_{+}$ of the 
generator $T$ in DS 
is given in Definition 5. Theorem 3 proves that the `dominating
parts' of $G_{+}$ are self-adjoint. In Theorem 7 it is proved that 
under General Assumption 1 and General Assumption 4, both, 
$G_{+}$ and $-G_{+}$ are generators of strongly continuous 
semigroups $T_{+}$ and $T_{-}$, resp. Theorem 13 shows the 
identity of the spectrum of $G_{+}$ with the spectrum of an 
operator polynomial generated by the operators $B$ and $A+C$.
Lemmas 14 and 15 show that certain simple rescalings 
of $A$ and $C$ which formally leave invariant (\ref{waveeq})
lead to theories which are related by a similarity transformation.  
Theorem 16 shows for a special case 
how these rescalings can be used 
to derive a better estimate for the growth of 
$T_{+}$ and  $T_{-}$ than the one induced 
by $(\ref{growth})$ in Lemma 6. Theorem 17 gives sufficient criteria
for stability in the form of inequalities which have to be satisfied
by the coefficients of the operator polynomial.
Part (ii) of this Theorem has been sucessfully applied in the 
discussion of the stability of the Kerr metric.
\cite{beyer4}
\newline
\linebreak
The rest of this section contains the mathematical details.   
\begin{ass}
In the following let $(X,<|>)$ be a non trivial complex Hilbert space. Denote 
by $\|\, \|$ the norm induced on $X$ by $<\,|\,>$. Further
let $A : D(A) \rightarrow X$ be
a densely defined linear self-adjoint operator in $X$ for 
which there is an $\varepsilon \in (0,\infty)$ such that
\begin{equation} \label{strictpos}
<\xi|A\xi> \,\, \geqslant \varepsilon <\xi|\xi>
\end{equation}
for all $\xi \in D(A)$. 
Denote by $A^{1/2}$ the 
square root of $A$ with domain $D(A^{1/2})$.
Further let be $B : D(A^{1/2}) \rightarrow X$ a linear 
operator in $X$ such that for some $a \in [0,1)$ and $b \in {\Bbb{R}}$ 
\begin{equation} \label{perturb}
\|B \xi \|^2 \leqslant a^2 \| A^{1/2} \xi \|^2 + b^2 \| \xi \|^2 
\end{equation} 
for all $\xi \in D(A^{1/2})$. 
Finally, let $C : D(A^{1/2}) \rightarrow X$ be linear 
and such that for some real numbers $c$ and $d$ 
\begin{equation} \label{perturbC}
\|C \xi \|^2 \leqslant c^2 \| A^{1/2} \xi \|^2 + d^2 \| \xi \|^2 
\end{equation} 
for all $\xi \in D(A^{1/2})$.
\end{ass}
Note that as a consequence of 
(\ref{strictpos})
the spectrum of $A$ is contained in the interval $[\varepsilon,\infty)$.
Hence $A$ is in particular positive and bijective 
and there is a uniquely defined linear and positive selfadjoint operator 
$A^{1/2}: D(A^{1/2}) \rightarrow X$ such that 
$(A^{1/2})^2 = A $. That operator is the so 
called {\it square root of} $A$.
Further note that from its definition and the bijectivity of $A$ 
follows that 
$A^{1/2}$ is in particular bijective. This can be concluded for instance as 
follows. By using the fact  
that $A^{1/2}$ commutes with $A$ 
it easy to see that for every
$\lambda \in [0, \varepsilon^{1/2})$ by 
$(A^{1/2}+\lambda) (A - \lambda^2)^{-1}$
there is given the inverse to $A^{1/2}-\lambda$.
Hence the spectrum of $A^{1/2}$ is contained in 
the interval $[\varepsilon^{1/2},\infty)$. All these facts will be used 
later on.
\begin{definition}
We define 
\begin{equation} \label{Y}
Y := D(A^{1/2}) \times X 
\end{equation}
and $(\, |\, ) : Y^2 \rightarrow {\Bbb{C}}$ by
\begin{equation}
(\xi | \eta) := <A^{1/2}\xi_1|A^{1/2}\eta_1> + <\xi_2|\eta_2>
\end{equation}
for all $ \xi = (\xi_1,\xi_2), \eta = (\eta_1,\eta_2) \in Y$. 
\end{definition}
Then we have the following 
\begin{theorem}
\begin{description}
\item[(i)]  $(Y,(\,|\,))$ is a complex Hilbert space.
\item[(ii)] The operator 
$H : D(A) \times D(A^{1/2}) \rightarrow Y$ in $Y$
defined by
\begin{equation}
 H \xi := (-i \xi_2, i A \xi_1)
\end{equation}
for all $ \xi = (\xi_1,\xi_2) \in  D(A) \times D(A^{1/2})$ is densely-defined,
linear and self-adjoint.
\item[(iii)] The operator $\hat{B} : D(H) \rightarrow Y$
defined by 
\begin{equation}
\hat{B} \xi := (0, -B \xi_2)
\end{equation} 
for all $ \xi = (\xi_1,\xi_2) \in  D(H)$ 
is linear. If $B$ is symmetric then 
$\hat{B}$ is symmetric, too. If $B$ is bounded
then $\hat{B}$ 
is bounded, too, and the corresponding operator norms
$\|B\|$ and  $|\hat{B}|$ satisfy
\begin{equation}  \label{estim}
|\hat{B}| \leqslant \|B\| \, \, .
\end{equation}
\item[(iv)]
The sum $H+\hat{B}$ is closed. If $B$ is symmetric 
then $H+\hat{B}$ is self-adjoint.
\item[(v)]
The operator $V : Y \rightarrow Y$ defined by
\begin{equation}
V \xi := (0,iC\xi_1)
\end{equation} 
for all $\xi = (\xi_1,\xi_2) \in Y$ is linear and bounded. The
operator norm $|V|$ of $V$ satisfies
\begin{equation}
|V| \leqslant ( c^2 + d^2 / \epsilon )^{1/2} \, \, .
\end{equation}
\end{description}
\end{theorem}
{\bfseries Proof}: (i): Obviously, $( \, | \,)$ defines a hermitean
sesquilinear form on $Y^2$. That $( \, | \,)$  is further positive definite 
follows from the positive definiteness of $<\,|\,>$ and 
the injectivity of $A^{1/2}$. Finally, the completeness of 
$(Y,|\,|)$, where $|\,|$ denotes the norm on $Y$ induced by $(\,|\,)$,
follows from the completeness of $(X,\|\,\|)$ together with the 
fact that $A^{1/2}$ has a {\it bounded} inverse.
Here 
it is essentially used that $0$ is not contained in  
the spectrum of $A$.
(ii): That $D(A) \times D(A^{1/2})$ is dense
in $Y$ is an obvious consequence of 
the facts that $D(A)$ is a core for $A^{1/2}$
(see e.g. Theorem 3.24 in chapter V.3 of \cite{kato})
and that $D(A^{1/2})$ is dense in $X$. The linearity of $H$ is obvious. Also 
the symmetry of $H$ follows straighforwardly 
from the symmetry of $A^{1/2}$. By that symmetry one gets further for any 
$\xi = (\xi_1,\xi_2) \in D(H^*)$ and any $\eta = (\eta_1,\eta_2) \in D(H)$:
\begin{eqnarray}
(H^* \xi | \eta ) &=& <\left(H^{*}\xi\right)_1 | A \eta_1> + 
                      <\left(H^{*}\xi\right)_2 | \eta_2 > \nonumber \\
= (\xi | H \eta ) &=&<-i \xi_2 | A \eta_1 > + 
                      <iA^{1/2} \xi_1 | A^{1/2} \eta_2 > 
\end{eqnarray}
and from this by using that $A$ is bijective and $A^{1/2}$ is self-adjoint 
that $\xi_1 \in D(A)$ and
\begin{equation}
\left(H^{*}\xi\right)_1 = -i \xi_2 \, \, , \, \, 
\left(H^{*}\xi\right)_2 = i A \xi_1 \, \, .
\end{equation}        
Hence $H$ is an extension of $H^{*}$ and thus $H = H^{*}$.  
(iii): The linearity of $\hat{B}$ is obvious. Also it is straightforward 
to see that $\hat{B}$ is symmetric if $B$ is symmetric. If $B$ is bounded
then
\begin{equation}
|\hat{B}\xi|^2 = \| B \xi_2 \|^2  \leqslant \| B \|^2 \|\xi_2\|^2 \leqslant
\| B \|^2 |\xi|^2
\end{equation}
for all $\xi = (\xi_1,\xi_2) \in D(H)$. Hence 
$\hat{B}$ is also bounded and 
$| \hat{B} |$, $\|B\|$ satisfy the claimed inequality.
(iv): Obviously, (\ref{perturb}) implies 
\begin{equation} \label{perturb1}
|\hat{B}\xi|^2 \leqslant a^2 |H \xi|^2 + b^2 |\xi|^2
\end{equation}
for all $\xi \in D(H)$. From this it is easily seen that 
$H+\hat{B}$ is closed (see, e.g., \cite{goldberg}, Lemma V.3.5).
Moreover in the case that $B$ (and hence by (iii)
also $\hat{B}$) is symmetric (\ref{perturb1}) implies according to the 
{\it Kato-Rellich} Theorem (see, e. g., Theorem X.12 in \cite{reedsimon} Vol. II) that $H + \hat{B}$ is self-adjoint. For the application of 
these theorems the assumption $a < 1$ made above is essential. 
(v) The linearity of $V$ is obvious. For any $\xi=(\xi_1,\xi_2) \in 
Y$ one has
\begin{eqnarray} \label{estimate}
| V\xi|^2 &=& \|i C \xi_1\|^2 \leqslant 
c^2 \| A^{1/2} \xi_1 \|^2 + d^2 \| \xi_1 \|^2 \nonumber \\
&=&
c^2 \| A^{1/2} \xi_1 \|^2 + d^2 \|(A^{1/2})^{-1}A^{1/2}\xi_1\|^2
\leqslant  
(c^2+d^2 / \epsilon) \, |\xi|^2 \, \, .
\end{eqnarray}
In the last step it has been used that
\begin{equation}
 \|(A^{1/2})^{-1}\| \leqslant 1/ \sqrt{\varepsilon} \, \, .
\end{equation}  
This follows by an application of the spectral theorem 
(see, e.g. Theorem VIII.5 in \cite{reedsimon} Vol. I) to $A^{1/2}$.
Since $\xi$ is otherwise arbitrary from (\ref{estimate}) follows
the boundedness of $V$ and the claimed inequality.$_\square$
\begin{ass}
In the following we assume in addition that $B$ is symmetric 
or bounded. 
\end{ass}
Note that condition (\ref{perturb}) is trivially satisfied if $B$ is bounded.
We define:
\begin{definition}
\begin{equation}
G_{+} := -i(H+\hat{B}+V) \, \, , \, \,   
G_{-} := i(H+\hat{B}+V) \, \, .
\end{equation}
\end{definition} 
then
\begin{lemma}
The operators $G_{+}$ and $G_{-}$ are closed and quasi-accretive. 
In particular 
\begin{equation} \label{growth}
Re (\xi| G \xi) \geqslant  - ( \mu_B  + |V|) \,  (\xi | \xi)  
\end{equation} 
for $G \in \{ G_{+},G_{-}\}$ and all $\xi \in D(H)$. 
Here $Re$ denotes the real part and
\begin{equation}
\mu_B :=     
 \left\{
 \begin{array}{cl}
 0 & \text{ if $B$ is symmetric} \\
 \|B\| & \text{if $B$ is bounded}
 \end{array}
 \right. \, \, .
\end{equation} 
\end{lemma}
{\bfseries Proof}: That $G_{+}$ and $G_{-}$ are closed is an obvious
consequence of (iv) and (v) of the previous theorem. Further if
$B$ is symmetric
one has because of (iv) and (v) of the preceeding theorem  
\begin{equation}
Re(\xi |G_{\pm}\xi) = \mp Re (\xi|iV\xi) \geqslant -|(\xi|iV\xi)|
\geqslant -|V| \, (\xi|\xi) 
\end{equation}
for all $\xi \in D(H)$.
Similarly, if $B$ is bounded one has because of (ii),(iii),(iv), 
(\ref{estim})
\begin{equation}
Re(\xi |G_{\pm} \xi) = \mp Re (\xi| i(\hat{B} + V) \xi) \geqslant 
-|(\xi|i(\hat{B}+V)\xi)| 
\geqslant -(\|B\|+|V|) \, (\xi|\xi) 
\end{equation} 
for all $\xi \in D(H)$. Hence in both cases $G_{+}$ and 
$G_{-}$ are 
quasi-accretive.$_\square$
\begin{theorem}
The operators $G_{+}$ and  
$G_{-}$ 
are infinitesimal generators of strongly continuous semigroups 
$T_{+} : [0,\infty) \rightarrow L(Y,Y)$ and 
$T_{-} : [0,\infty) \rightarrow L(Y,Y)$, respectively. 
If $\mu_{\pm} \in {\Bbb{R}}$ are such that 
\begin{equation}
Re(\xi|G_{\pm}\xi) \geqslant - \mu_{\pm} \, (\xi|\xi)
\end{equation}
for all $\xi \in D(H)$ 
the spectra
of $G_{+}$ and $G_{-}$ are contained in the half-plane 
$[- \mu_{+} ,\infty) \times {\Bbb{R}}$ and 
$[- \mu_{-} ,\infty) \times {\Bbb{R}}$, respectively, and  
\begin{equation}
|T_{+}(t)| \leqslant \exp (\mu_{+} t) \, \, , \, \, 
|T_{-}(t)| \leqslant \exp (\mu_{-} t)
\end{equation} 
for all $t \in [0,\infty)$.
\end{theorem}
{\bfseries Proof}:
Obviously, by the Lumer-Phillips theorem 
(see, e.g., Theorem X.48 in Vol. II of \cite{reedsimon})
and the preceeding lemma the 
theorem follows if we can show that there is a real number $\lambda
< \min\{-\mu_{+},-\mu_{-}\}$
such that $G_{\pm} - \lambda$ has a dense range in $Y$. 
For that proof let be $\xi$ some element of $D(H)$ and $\lambda$ any
real number such that $|\lambda| \geqslant |V|^2$. 
Then we get from the symmetry of $H$  
\begin{equation}
|(H-i\lambda)\xi|^2 = |H\xi|^2 + \lambda^2 |\xi|^2
\end{equation}
and
\begin{equation}
|(H-i\lambda)\xi| \geqslant \max\{ |H\xi|, 
|\lambda|^{1/2} |V\xi|\} \, \, .
\end{equation}
Using these identities together with (\ref{perturb}) 
\begin{eqnarray}
|(\hat{B}+V)\xi |^2 &\leqslant& 
|\hat{B} \xi|^2 + 2|\hat{B}\xi| \, |V\xi|+ |V\xi|^2 \nonumber \\
&\leqslant&  a^2 |H \xi|^2 + 
2|\hat{B}\xi| \, |V\xi| + (b^2 + |V|^2) |\xi|^2
\nonumber \\ 
&\leqslant&  a^2 |H \xi|^2 + 2a|H\xi| \, |V\xi| + (b + |V|)^2 |\xi|^2 
\nonumber \\
&\leqslant&  a^2 |(H-i\lambda) \xi|^2 + 
2a|(H-i \lambda)\xi| \, |V\xi| + [(b + |V|)^2 - a^2 \lambda^2] |\xi|^2 
\nonumber \\
&\leqslant& 
a(a+2|\lambda|^{-1/2}) |(H-i\lambda) \xi|^2 + 
[(b + |V|)^2 - a^2 \lambda^2] |\xi|^2 \, \, .
\end{eqnarray}
Hence for any real $\lambda $ with
\begin{equation}
|\lambda| > \max\{|V|^2,4(1-a)^{-2},(b+|V|)/a, |\mu_{+}|,|\mu_{-}|\} \, \, ,
\end{equation} 
where we assume without restriction that $a > 0$, we get
\begin{equation}
|(\hat{B}+V)\xi | \leqslant a^{\prime} |(H-i\lambda) \xi|
\end{equation}
where $a^{\prime}$ is some real number from $[0,1)$.
Since $\xi \in D(H)$ is otherwise arbitrary, we conclude that
\begin{equation}
(\hat{B} + V)(H-i\lambda)^{-1}
\end{equation}
defines a bounded linear operator on $Y$ with operator norm
smaller than $1$. 
Since
\begin{equation}
H+ \hat{B} + V -i\lambda =
\left( 1 + (\hat{B} + V)(H-i\lambda)^{-1} \right) (H-i\lambda)
\end{equation}
we conclude that 
$H + \hat{B} + V -i\lambda$ is bijectice and hence also 
that $G_{+} - \lambda$ and $G_{-} - \lambda$ are both bijective.
Hence the theorem follows.$_{\square}$
\newline \linebreak
We note that General Assumption 4 has been used only to conclude 
that $G_{+}$ and  $G_{-}$ are {\it both} quasi-accretive. Now it is 
easy to see that if $B$ is in addition such that 
$iB$ is quasi-accretive (but not necessarily bounded or 
antisymmetric)
then $-i\hat{B}$ and hence also   
$G_{+}$ are quasi-accretive, too. 
As a consequence
we have the following 
\begin{corollary}
Instead of General Assumption 4 let $B$ be such that 
$iB$ is quasi-accretive. Then $G_{+}$ 
is the infinitesimal generator of a strongly continuous semigroup 
$T_{+} : [0,\infty) \rightarrow L(Y,Y)$.
If $\mu_{+} \in {\Bbb{R}}$ is such that 
\begin{equation} \label{stability}
Re(\xi|G_{+}\xi) \geqslant - \mu_{+} \, (\xi|\xi)
\end{equation}
for all $\xi \in D(H)$ 
the spectrum
of $G_{+}$ is contained in the half-plane 
$[- \mu_{+} ,\infty) \times {\Bbb{R}}$ and  
\begin{equation}
|T_{+}(t)| \leqslant \exp (\mu_{+} t) 
\end{equation} 
for all $t \in [0,\infty)$.
\end{corollary}
Theorem 7 has the following  
\begin{corollary}
\begin{description}
\item[(i)]
By 
\begin{equation}
T(t) :=     
 \left\{
 \begin{array}{cl}
 T_{+}(t) & \text{for $t \geqslant 0$} \\
 T_{-}(-t) & \text{for $t < 0$}
 \end{array}
 \right. 
\end{equation} 
for all $t \in {\Bbb{R}}$
there is defined a strongly continuous group $T : {\Bbb{R}} \rightarrow
L(Y,Y)$.
\item[(ii)] 
For every $t_0 \in {\Bbb{R}}$ and every $\xi \in D(G_{+})$ 
there is a uniquely determined 
differentiable map $u : {\Bbb{R}} \rightarrow Y$
such that 
\begin{equation}
u(t_0) = \xi
\end{equation}
and 
\begin{equation} \label{eveq}
u^{\prime}(t) = - G_{+} u(t) 
\end{equation}
for all $t \in {\Bbb{R}}$. Here $\, ^{\prime}$ denotes differentiation 
of functions assuming values in $Y$.
\item[(iii)]
The function $(u|u) : {\Bbb{R}} \rightarrow {\Bbb{R}}$ defined by 
\begin{equation}
(u|u)(t) := (\,u(t)|u(t)\,) \, \, , t \in {\Bbb{R}}
\end{equation}
is differentiable and
\begin{equation}
(u|u)^{\prime}(t) = - 2 Re\,(\,u(t)|G_{+}u(t)\,)
\end{equation}
for all $t \in {\Bbb{R}}$.
\end{description} 
\end{corollary}
{\bfseries Proof}: The corollary follows from Theorem 7 by 
standard results of semigroup theory. For instance, see
section 1.6 in \cite{pazy} for (i) and section 
IX.3 in \cite{kato} for (ii). (iii) is an obvious consequence of (ii).
$_\square$
\newline
\linebreak
Note in particular the {\it the special case}  
\footnote{Such cases
are easy to construct.}
that there is a non trivial element 
$\eta$ in the kernel of $A+C$ for which there is 
$\xi \in D(A)$ such that
\begin{equation}
(A+C)\xi = -iB\eta \, \, .
\end{equation}
Then by
\begin{equation}
u(t) := (\xi + t \eta , \eta) \, \, , \, \, t \in {\Bbb{R}}
\end{equation}
there is given a growing solution of (\ref{eveq}).
\newline
\linebreak
The following lemma is needed in the formulation of the subsequent 
theorem. 
\begin{lemma}
By 
\begin{equation}
\| \xi \|_{A^{1/2}} := \|A^{1/2}\xi\| \, \, , \, \, \xi \in D(A^{1/2})
\end{equation}
there is defined a norm $\| \, \|_{A^{1/2}}$ on $D(A^{1/2})$. 
Moreover 
\begin{equation}
W_1 := (D(A^{1/2}),\| \, \|_{A^{1/2}})
\end{equation}
is complete.
\end{lemma}
{\bfseries Proof}: The lemma is a trivial consequence of
the completeness of $X$ and the bijectivity of $A^{1/2}$.$_\square$
\begin{theorem}
Let be $t_0 \in {\Bbb{R}}$, $\xi \in D(A)$ and 
$\eta \in D(A^{1/2})$. Then there is a uniquely determined 
differentiable map
$u : {\Bbb{R}} \rightarrow W_1$
with 
\begin{equation} \label{initcond}
u(t_0) = \xi \, \, \text{and} \, \, u^{\,\prime}(t_0) = \eta 
\end{equation}
and such 
that $u^{\, \prime} : {\Bbb{R}} \rightarrow X$ is differentiable with
\begin{equation} \label{waveeq}
(u^{\,\prime})^{\,\prime}(t) + iB u^{\,\prime}(t) + (A+C)u(t) = 0 
\end{equation}
for all $t \in {\Bbb{R}}$.
\end{theorem}
{\bfseries Proof}: For this let be $v=(v_1,v_2):{\Bbb{R}} \rightarrow Y$ be 
such that 
\begin{equation} \label{in}
v(t_0) = (\xi,\eta)
\end{equation}
and 
\begin{equation} \label{eq}
v^{\, \prime}(t) = -G_{+} v(t) \, \, , t \in {\Bbb{R}} \, \, .
\end{equation}
Such $v$ exists according to 
Corollary 9 (ii).
Using the continuity of the canonical
projections of $Y$ onto $W_1$ and $X$
it is easy to see that $u := v_1$
is a differentiable map into $W_1$ such that $u^{\, \prime} : {\Bbb{R}}
\rightarrow X$ is differentiable and such that (\ref{initcond}),
(\ref{waveeq}) are both satisfied. On the other hand if $u : {\Bbb{R}}
\rightarrow W_1$ has the properties stated in the corollary    
it follows by the continuity of the canonical imbeddings
of $W_1$, $X$ into $Y$ that 
$w := (u,u^{\, \prime})$ satisfies both equations (\ref{in})
and (\ref{eq}). Then $u = v_1$ follows by Corollary 9 (ii). 
$_\square$ 
\begin{corollary}
In addition to the assumptions made let $C$ be in particular 
bounded. \footnote{Note that
in this case (\ref{perturbC}) is trivially satisfied.}
Further let 
$u : {\Bbb{R}} \rightarrow W_1$ be differentiable with a differentiable 
derivative  $u^{\, \prime} : {\Bbb{R}} \rightarrow X$
and such that (\ref{waveeq}) holds. Finally, define $E_{u} : {\Bbb{R}} \rightarrow
{\Bbb{R}}$ by 
\begin{equation} \label{energy}
E_u(t) := \frac{1}{2} 
\left( <u^{\, \prime}(t)|u^{\, \prime}(t)> + <u(t)|(A+Re(C))u(t)>  \right) \, \, .
\end{equation}
Then $E_u$ is differentiable and 
\begin{eqnarray} \label{energyconserv}
E_u^{\, \prime}(t) = 
\left\{
 \begin{array}{cl}
 - Im <u(t)|Im(C) u^{\,\prime}(t)> & \text{for symmetric $B$} \\
  \frac{1}{2} <u^{\,\prime}(t)|Im(B) u^{\,\prime}(t)> 
- Im <u(t)|Im(C) u^{\,\prime}(t)> & \text{for bounded $B$}
 \end{array}
 \right.
 \end{eqnarray}
for all $t \in {\Bbb{R}}$,
where for any bounded linear operator $F$ on $X$:
\begin{equation}
Re(F) := \frac{1}{2} \left( F + F^* \right) \, \, , \, \, 
Im(F) := \frac{1}{2i} \left( F - F^* \right) \, \, .
\end{equation}
\end{corollary}
{\bfseries Proof}: For this define $v := (u,u^{\, \prime})$. Then according
to the preceeding proof $v$ satisfies (\ref{eq}). { \it For a 
symmetric} $B$  it follows by Corollary 9 and 
Theorem 3 (iv) that 
\begin{eqnarray} \label{conserv}
(v|v)^{\prime}(t) &=& 2 Re\,(\,v(t)|iV v(t)\,) \nonumber \\
&=& -<u^{\, \prime}(t)| Cu(t)> -<Cu(t)|u^{\, \prime}(t)> \\
&=& -<u|Re(C)u>^{\, \prime}(t) -2 Im <u(t)|Im(C) u^{\,\prime}(t)> \nonumber
\end{eqnarray}
for all $t \in {\Bbb{R}}$. In the last step it has been used that 
$u$ is also differentiable with the same derivative 
viewed as map with values in $X$. This 
follows from the fact the canonical imbedding of $W_1$ into $X$ 
is continuous since $A^{1/2}$ is bijective. Further the definition 
\begin{equation}
<u|Re(C)u>(t) := <u(t)|Re(C)u(t)> \, \, , \, \, t \in {\Bbb{R}} 
\end{equation}
for the map $<u|Re(C)u> : {\Bbb{R}} \rightarrow {\Bbb{R}} $ has been used.
Obviously, (\ref{energyconserv})
follows from (\ref{conserv}) by using definition (\ref{energy}). 
In this step also the symmetry of $A^{1/2}$ 
is used together with the fact that $u$ assumes values in $D(A)$.
{ \it For a 
bounded} $B$ by Corollary 9 and 
Theorem 3 (ii) follows that
\begin{eqnarray} \label{conserv2}
(v|v)^{\prime}(t) = &&2 Re\,(\,v(t)|i(\hat{B}+V) v(t)\,) \nonumber \\
=&&2Im<u^{\, \prime}(t)|Bu^{\, \prime}(t)> \nonumber \\
&&-<u^{\, \prime}(t)| Cu(t)> -<Cu(t)|u^{\, \prime}(t)> \\
=&&2<u^{\, \prime}(t)|Im(B)u^{\, \prime}(t)> \nonumber \\
&&-<u|Re(C)u>^{\, \prime}(t) -2 Im <u(t)|Im(C) u^{\,\prime}(t)> \nonumber
\end{eqnarray}
for all $t \in {\Bbb{R}}$. 
Obviously, (\ref{energyconserv})
follows from (\ref{conserv}) by using definition (\ref{energy}).
$_{\square}$
\newline
\linebreak
The next theorem relates the spectrum of $G_+$ to the spectrum 
of the so called {\it operator polynomial} $A + C - \lambda B -\lambda^2 $,
where $\lambda$ runs through the complex numbers. \cite{markus,rodman}
\begin{theorem} Let $\lambda$ be some complex number. 
\begin{description} 
\item[(i)] Then  
$H+\hat{B}+V - \lambda$ is not injective if and only if
$A + C - \lambda B -\lambda^2 $ is not injective. If 
$H+\hat{B}+V - \lambda$ is not injective then
\begin{equation}
\ker(H+\hat{B}+V - \lambda) = 
\{ (\xi, i \lambda \xi) : \xi \in 
\ker( A + C - \lambda B -\lambda^2 ) \} \, \, .
\end{equation} 
\item[(ii)]
Further 
$H+\hat{B}+V - \lambda$ is bijective if and only if
$A + C - \lambda B -\lambda^2 $ is bijective. 
If $H+\hat{B}+V - \lambda$ is bijective then 
for all
$\eta = ( \eta_1,\eta_2) \in Y$:
\begin{equation}
( H+\hat{B} + V - \lambda)^{-1} \eta = 
\left(\xi,i(\lambda \xi + \eta_1)\right) \, \, ,
\end{equation}
where
\begin{equation} \label{resolv}
\xi = ( A + C - \lambda B -\lambda^2 )^{-1} 
[(B+\lambda)\eta_1 - i \eta_2] \, \, .
\end{equation}
\end{description}
\end{theorem} 
{\bfseries Proof}: (i) If $H+\hat{B}+V - \lambda$ is not injective
and $ \xi = (\xi_1,\xi_2) \in \ker(H+\hat{B}+V - \lambda)$ it follows
from the definitions in theorem 3 that 
\begin{equation}
\xi_2 = i \lambda \xi_1 \, \, , \, \, 
(A + C - \lambda B -\lambda^2)\xi_1 = 0  
\end{equation}
and hence also that 
$A + C - \lambda B -\lambda^2$ is not injective. 
If $A + C - \lambda B -\lambda^2 $ is not injective it follows again from 
the definitions in theorem 3 that
\begin{equation}
(H+\hat{B}+V - \lambda)(\xi,i \lambda \xi) = 0
\end{equation}
and hence also that 
$H+\hat{B}+V - \lambda$ is not injective.
(ii) If $H+\hat{B}+V - \lambda$ is bijective it follows by (i) that 
$A + C - \lambda B -\lambda^2$ is injective. For $\eta \in X$
and $\xi = (\xi_1,\xi_2) := 
(H+\hat{B}+V - \lambda)^{-1}(0, i \eta)$ 
it follows from the definitions in theorem 3 that
\begin{equation}
(A + C - \lambda B -\lambda^2)\xi_1 = \eta
\end{equation} 
and hence that $A + C - \lambda B -\lambda^2$ is also surjective.
If $A + C - \lambda B -\lambda^2$ is bijective it follows by (i)
that $H+\hat{B}+V - \lambda$ is injective. Further if
$\eta = (\eta_1,\eta_2) \in Y$ and $\xi$ is defined by (\ref{resolv})
it follows from the definitions in theorem 3 that 
\begin{equation}
(H+\hat{B}+V - \lambda)(\xi,i(\lambda \xi + \eta_1)) = \eta
\end{equation} 
and hence that $H+\hat{B}+V - \lambda$ is also surjective.$_\square$
\begin{lemma}
Let be $\varepsilon^{\, \prime} < \varepsilon$ and 
\begin{equation} \label{prime}
A^{\, \prime} := A - \varepsilon^{\, \prime} \, \, , \, \,
C^{\, \prime} := C + \varepsilon^{\, \prime} \, \, .
\end{equation}
Then
\begin{description}
\item[(i)] 
\begin{equation}  \label{id1}
D(A^{\,\prime 1/2}) = D(A^{1/2})
\end{equation}
and for all $\xi \in D(A^{1/2})$ 
\begin{equation}  \label{id2}
\|A^{1/2}\xi\|^2 = \|A^{\, \prime 1/2 \,}\xi\|^2 + 
\varepsilon^{\, \prime} \|\xi\|^2 \, \, . 
\end{equation}
\item[(ii)]
The operators $A^{\, \prime}, B$ and $C^{\, \prime}$ satisfy
\begin{eqnarray}
<\xi|A^{\, \prime} \xi> \,\, \quad \geqslant &&
(\varepsilon - \varepsilon^{\, \prime}) <\xi|\xi> \nonumber \\
\|B \xi \|^2 \quad \leqslant && a^2 \| A^{\, \prime 1/2} \xi \|^2 + 
(a^2 \varepsilon^{\, \prime} + b^2) \| \xi \|^2  \nonumber \\
\|C^{\, \prime} \xi \|^2 \quad \leqslant &&
|c| \left[\, |c| +  2 |\varepsilon^{\, \prime}| 
\, (\varepsilon - \varepsilon^{\, \prime})^{-1/2} 
\right] \| A^{\, \prime \, 1/2} \xi \|^2 + \nonumber \\
&& \left[\, |\varepsilon^{\, \prime}| +
( c^2 |\varepsilon^{\, \prime}| + d^2)^{1/2} \, \right]^2 
\, \| \xi \|^2  
\end{eqnarray}
for all $\xi \in D(A^{1/2})$. 
\end{description}
\end{lemma}
{\bfseries Proof}: (i) First, since 
$\varepsilon^{\, \prime} < \varepsilon$
by (\ref{prime}) there is defined a linear self-adjoint and 
positive operator $A^{\, \prime}$ in $X$.
Obviously, using the symmetry of $A^{1/2}$ and $A^{\, \prime 1/2}$ 
(\ref{id2}) follows for 
all elements of $D(A)$.  From this 
(\ref{id1}) and (\ref{id2}) follow straightforwardly 
by using the facts that $D(A)$ is a core for 
both, $A^{1/2}$ and $A^{\, \prime 1/2}$ 
(see e.g. Theorem 3.24 in chapter V.3 of \cite{kato}),
that $X$ is complete and that  
both operators, $A^{1/2}$ and $A^{\, \prime 1/2}$
are closed. (ii) The first two inequalities 
are obvious consequences of the corresponding ones in 
General Assumption 1, the definition (\ref{prime}) and
of (\ref{id2}). For the proof of the third 
we notice that from the first inequality along with 
an application of the spectral theorem 
(see, e.g. Theorem VIII.5 in \cite{reedsimon} Vol. I) to 
$A^{\, \prime \, 1/2}$ follows that 
\begin{equation} \label{spectral}
 \|(A^{\, \prime \, 1/2})^{-1}\| \leqslant 1/ 
\sqrt{\varepsilon - \varepsilon^{\, \prime}} \, \, .
\end{equation}  
Further from General Assumption 1
and (\ref{id2}) one gets
\begin{equation} 
\|C \xi \|^2 \leqslant c^2 \| A^{\, \prime \, 1/2} \xi \|^2 + 
(c^2 |\varepsilon^{\, \prime}| + d^2) \| \xi \|^2 \, \, . 
\end{equation}
for all $\xi \in D(A^{1/2})$.
From these inequalities we get
\begin{eqnarray}
\|C^{\, \prime} \xi \|^2 &\leqslant& 
\|C \xi\|^2 + 2 | \varepsilon^{\, \prime}| \, \|C \xi\| \,
\| \xi \|+ \varepsilon^{\, \prime\, 2} \| \xi \| ^2 \\
&\leqslant&
c^2 \| A^{\, \prime \, 1/2} \xi \|^2 + 
(\varepsilon^{\, \prime \, 2} + 
c^2 |\varepsilon^{\, \prime}| + d^2) \| \xi \|^2 +
2 | \varepsilon^{\, \prime}| \, \|C \xi\| \,
\| \xi \|  
\nonumber \\
&\leqslant&
c^2 \| A^{\, \prime \, 1/2} \xi \|^2 + 
\left[\, |\varepsilon^{\, \prime}| +
( c^2 |\varepsilon^{\, \prime}| + d^2)^{1/2} \, \right]^2 
\, \| \xi \|^2 +
2 | \varepsilon^{\, \prime}| \, |c| \, \|A^{\, \prime \, 1/2} 
\xi\| \,
\| \xi \| 
\nonumber \\ 
&\leqslant& 
|c| \left[\, |c| +  2 |\varepsilon^{\, \prime}| 
\, (\varepsilon - \varepsilon^{\, \prime})^{-1/2} 
\right] \| A^{\, \prime \, 1/2} \xi \|^2 + 
\left[\, |\varepsilon^{\, \prime}| +
( c^2 |\varepsilon^{\, \prime}| + d^2)^{1/2} \, \right]^2 
\, \| \xi \|^2  \nonumber
\end{eqnarray}
for all $\xi \in D(A^{1/2})$
and hence the third inequality.$_\square$ 
\newline
\linebreak
As a consequence of (ii) the sequence
$X,A^{\, \prime},B,C^{\, \prime}$ satisfies 
General Assumption 1. The corresponding $Y$ given by 
Definition 2 is because of (i) again given by (\ref{Y}).
Moreover the corresponding norm $|\,|^{\, \prime}$
on $Y$ turns out to be equivalent to 
$|\,|$. More precisely one has for every 
$\varepsilon^{\, \prime} \leqslant 0 $
\begin{lemma}
\begin{equation}
|\,\,|^{\, \prime} \leqslant | \,\, |
\leqslant \varepsilon^{1/2} \,
(\varepsilon - \varepsilon^{\, \prime})^{-1/2} \, 
|\,\,|^{\, \prime} 
\end{equation}
and for every bounded linear operator $F$ on $Y$:
\begin{equation}
\varepsilon^{-1/2} \,
(\varepsilon - \varepsilon^{\, \prime})^{1/2} \,
|F|^{\, \prime} \leqslant |F|
\leqslant 
\varepsilon^{1/2} \,
(\varepsilon - \varepsilon^{\, \prime})^{-1/2} \,
|F|^{\, \prime} \, \, .
\end{equation}
\end{lemma}
{\bfseries Proof}: The first inequality is a straightforward 
consequence of (\ref{id2}) and (\ref{spectral}). The second 
inequality is a straightforward implication of the first. 
$_\square$.
\newline
\linebreak
Note that the $G_{\pm}$ corresponding to the 
the sequence $X,A^{\, \prime},B,C^{\, \prime}$ 
are the same for all $\varepsilon^{\, \prime}$
($\varepsilon^{\, \prime}$ drops out of the definition). 
Moreover as a consequence of the preceding lemma the
the topologies induced on $Y$ are equivalent. Hence 
the generated groups are the same, too. 
This will be used in the following important 
special case.
\begin{theorem}
Let be $A = A_0 + \varepsilon$, where $A_0$ is a 
densely defined linear positive self-adjoint operator 
and let be $C = - \varepsilon$. Then
\begin{equation} \label{estgrowth}
|T_{\pm}(t)| \leqslant e \, \varepsilon^{1/2} \, t \, \exp(\mu_B t)
\end{equation}
for all $t \geqslant \varepsilon^{-1/2}$.
\end{theorem}
{\bfseries Proof}: For this let be 
$\varepsilon^{\, \prime} \in [0, \varepsilon)$
and define $A^{\, \prime}$ and $C^{\, \prime}$
as in Lemma 14. Hence  
\begin{equation}
A^{\, \prime} = A_0 + \varepsilon -\varepsilon^{\, \prime}
\, \, , \, \,  
C^{\, \prime} = - (\varepsilon -\varepsilon^{\, \prime}) \, \, .
\end{equation}
Then from Theorem 3(v), Lemma 6 and Theorem 7 we conclude that
\begin{equation}
|T_{\pm}(t)|^{\, \prime} 
\leqslant \exp\left(\, [\mu_B+
(\varepsilon -\varepsilon^{\, \prime})^{1/2}] \, t \, \right)
\end{equation}
for all $t \in {\Bbb{R}}$ and hence by 
Lemma 15 that
\begin{equation}
|T_{\pm}(t)|
\leqslant
\varepsilon^{1/2} \,
(\varepsilon - \varepsilon^{\, \prime})^{-1/2} \,
\exp\left(\, [\mu_B+
(\varepsilon -\varepsilon^{\, \prime})^{1/2}] \, t \, \right) \, \, .
\end{equation} 
For $t \geqslant \varepsilon^{-1/2}$ we get from this 
(\ref{estgrowth}) by choosing  
\begin{equation}
\varepsilon^{\, \prime} := 
\varepsilon - t^{-2} \, \, ._{\square}
\end{equation}
Note that in this special case (\ref{energy}) is 
conserved and positive. 
\newline
\linebreak
We are now giving stability criteria.
\begin{theorem} In addition let $B$ and $C$ be both symmetric.
\begin{description} 
\item[  (i)]
Let $A$, $B$ and $C$ be such that
\begin{equation} \label{crit1}
<\xi|(A+C)\xi> + \frac{1}{4} <\xi|B\xi>^2 \, \, \geqslant 0
\end{equation}
for all $\xi \in D(A)$ with $\|\xi\|=1$.
Then the spectrum of $iG_{+}$ is real.
\item[(ii)]
In addition let $B$ and $C$ be both bounded
and let $A+C + (b/2)B -(b^2/4)$ be
positive for some $b \in {\mathbb{R}}$.
Then the spectrum of $iG_{+}$ is real and there are $K\geqslant 0$ and 
$t_0 \geqslant 0$ such that 
\begin{equation} \label{growthT}
|T(t)| \leqslant K|t|  
\end{equation}
for all $|t| \geqslant t_0$.
\end{description}
\end{theorem}
{\bfseries Proof}: (i): 
First from General Assumption 1 and the assumed symmetry of $B$ and $C$ 
follows that, both, by $A^{-1/2}BA^{-1/2}$ and $A^{-1/2}CA^{-1/2}$
there is given a bounded symmetric and hence (by the theorem of 
Hellinger and Toplitz) also self-adjoint operator on $X$. Hence 
\begin{equation} \label{defAlambda}
A(\lambda) :=  \lambda^2 A^{-1} +
\lambda A^{-1/2}BA^{-1/2}
-\left(
1+A^{-1/2}CA^{-1/2} \right) 
\, \, , \, \, 
\lambda \in {\mathbb{C}}
\end{equation}
defines a self-adjoint operator polynomial in $L(X,X)$. 
In addition one has 
$A^{-1} \geqslant 1 / \varepsilon$ . 
Further for every $\xi \in D(A^{1/2})$ and $\lambda \in {\mathbb{C}}$ 
\begin{equation} \label{identity1}
<\xi|A(\lambda)\xi> = <\eta|A^{1/2}A(\lambda)A^{1/2}\eta> =
- <\eta | (A+C - \lambda B - \lambda^2) \eta>
\end{equation}
where $\eta := A^{-1/2} \xi \in D(A)$.
Now (\ref{crit1}) implies that the roots of the polynomial
$<\eta|(A + C - \lambda B - \lambda^2)\eta> , \lambda \in {\mathbb{C}}$  
are real. Hence by (\ref{identity1})
the roots of $<\xi|A(\lambda)\xi> , \lambda \in {\mathbb{C}}$ 
are real, too. Since $\xi \in D(A^{1/2})$
is otherwise arbitrary and  $D(A^{1/2})$ is dense in 
$X$ this implies also that $<\xi|A(\lambda)\xi>$
has only real roots for all $\xi \in X$. Hence (see \cite{markus}, 
Lemma 31.1) the polynomial $A(\lambda) , \lambda \in {\mathbb{C}}$ is 
weakly hyperbolic and has therefore a real spectrum.
As a consequence $A(\lambda)$ is bijective for all non real 
$\lambda$. Now for any such $\lambda$
\begin{equation} \label{identity2}
A+C - \lambda B - \lambda^2 = 
 - A^{1/2}A_{\bracevert}(\lambda)A^{1/2}_{\bracevert} \, \, , 
\end{equation}  
where $A^{1/2}_{\bracevert}$ denotes the restriction of $A^{1/2}$,
both, to $D(A)$ in domain and $D(A^{1/2})$ in range and
$A_{\bracevert}(\lambda)$ denotes the restriction of $A(\lambda)$
to $D(A^{1/2})$, both, in domain and in range. For this note that
$A(\lambda)$ leaves $D(A^{1/2})$ 
invariant. Further
from the bijectivity of $A^{1/2}$, 
$A(\lambda)$ and  (\ref{defAlambda}) 
follows the bijectivity of $A^{1/2}_{\bracevert}$ 
and $A_{\bracevert}(\lambda)$, respectively and hence 
by (\ref{identity2}) that $A+C - \lambda B - \lambda^2$ is 
bijective. This is true for all non real $\lambda$ 
and hence it follows by Theorem 13 that the spectrum of $iG_{+}$ is real. 
(ii) So let $B$ and $C$ be both bounded
and let $A+C + (b/2)B -(b^2/4)$ be
positive for some $b \in {\mathbb{R}}$.
In addition let be $\epsilon$ some 
real number greater than zero and define 
\begin{equation}
A^{\, \prime} := A+C+(b/2)B-(b^2/4) + \varepsilon \, \, , \, \, 
C^{\, \prime} := - \varepsilon \, \, , \, \,
B^{\, \prime} := B - b \, \, .
\end{equation}
First it is observed that
\begin{equation} \label{dom}
 D(A^{\, \prime \, 1/2}) = D(A^{1/2})
\end{equation} 
and that there exist nonvanishing real constants 
$K_1$ and  $K_2$ such that
\begin{equation} \label{equiv}
K_1^2 \| A^{1/2} \xi \|^2 \leqslant 
\| A^{\, \prime \, 1/2} \xi \|^2 \leqslant
K_2^2 \| A^{1/2} \xi \|^2 
\end{equation}
for every $\xi \in D(A^{1/2})$.
This can be proved as follows. Obviously, by the symmetry of 
$A^{1/2}$ and $A^{\, \prime \, 1/2}$, 
the Cauchy-Schwarz inequality, the boundedness of $B$, $C$,
$A^{-1/2}$ and  $A^{\, \prime \, -1/2}$ follows the existence of  
nonvanishing real constants 
$K_1$ and  $K_2$ such that (\ref{equiv}) is valid for 
all $\xi \in D(A)$. Since $D(A)$ is a core for both,
$A^{1/2}$ and $A^{\, \prime \, 1/2}$ 
(see e.g. Theorem 3.24 in chapter V.3 of \cite{kato})
from that inequality follows (\ref{dom}) and (\ref{equiv})
for all $\xi \in D(A^{1/2})$. Note that to conclude this it is used
that $X$ is complete and that, both, $A^{1/2}$ and 
$A^{\, \prime \, 1/2}$ are closed.
\newline 
\linebreak
Obviously, from the assumptions made follows that
also $A^{\, \prime}, B^{\, \prime}$ and $C^{\, \prime}$ 
instead of $A$, $B$ and $C$, respectively, satisfy 
General Assumption 1 and General Assumption 4. Hence
by Theorem 16 follows that
\begin{equation} \label{growthTpm}
|T_{\pm}^{\, \prime}(t)|^{\, \prime} 
\leqslant e \, \varepsilon^{1/2} \, t 
\end{equation}
for all $t \geqslant \varepsilon^{-1/2}$, where primes
indicate quantities whose definition uses one or more of
the operators $A^{\, \prime}, B^{\, \prime}$ and $C^{\, \prime}$
instead of  $A, B$ and $C$.
In addition (\ref{dom}) and (\ref{equiv})
imply $Y = Y^{\, \prime}$ as well as the equivalence of the norms 
$|\,|$ and  $|\,|^{\, \prime}$. Now define the auxiliar 
transformation $S_0:Y^{\, \prime} \rightarrow Y$ by
\begin{equation}
S_0\xi := \left(\xi_1, \xi_2 - i(b/2)\xi_1\right) 
\end{equation}
for all $\xi = (\xi_1,\xi_2) \in Y^{\, \prime}$. Obviously, $S_0$
is bijective and bounded with the bounded inverse $S_{0}^{-1}$ given by 
$S_0^{-1}\xi := \left(\xi_1, \xi_2 + i(b/2)\xi_1\right)$ for
all $\xi = (\xi_1,\xi_2) \in Y$. In addition define 
$S_{\pm} : [0,\infty) \rightarrow L(Y,Y)$ by
\begin{equation} \label{relST}
S_{\pm}(t) := \exp(\mp ibt/2) S_{0}T^{\, \prime}_{\pm}(t) S^{-1}_{0} \, \, ,
\end{equation}
for all $t \in [0,\infty)$. Obviously, $S_{\pm}$ defines
a strongly continuous semigroup with the corresponding generator
\begin{equation}
S_{0}\, G^{\, \prime}_{\pm}\,S^{-1}_{0} \pm i \frac{b}{2} = G_{\pm} 
\, \, . 
\end{equation} 
This implies $S_{\pm} = T_{\pm}$ and by (\ref{growthTpm})
and (\ref{relST}) the existence of $K\geqslant 0$ and 
$t_0 \geqslant 0$ such that (\ref{growthT}) is valid 
for all $|t| \geqslant t_0$. Finally, from this follows by the Theorem
of Hille-Yosida-Phillips that the spectrum of $iG_{+}$ is real.$_\square$
\begin{lemma}
Let $D$ be a core for $A$. 
Further let be $B_0 : D \rightarrow X$ a linear 
operator in $X$ such that for some 
real numbers $a_0$ and $b_0$ 
\begin{equation} \label{perturbB0}
\|B_0 \xi \|^2 \leqslant a_0^2 
<\xi|A\xi> + \, b_0^2 \, \| \xi \|^2 
\end{equation} 
for all $\xi \in D$. 
Then there is a uniquely determined
linear extension $\bar{B}_0 : D(A^{1/2}) \rightarrow X$
of $B_0$ such that
\begin{equation} \label{perturbbarB0}
\|\bar{B}_0 \xi \|^2 
\leqslant a_0^2 \, \| A^{1/2} \xi \|^2 + b_0^2 \, \| \xi \|^2 
\end{equation}
for all $\xi \in D(A^{1/2})$. If $B_0$ is in addition
symmetric $\bar{B}_0$ is symmetric, too. 
\end{lemma}
{\bfseries Proof}: First we notice that 
$D$ is a core for $A^{1/2}$, too. Obviously, since   
$D(A)$ is a core for 
$A^{1/2}$ (see e.g. Theorem 3.24 in chapter V.3 of \cite{kato})
this follows if we can show that the closure of
the restriction of $A^{1/2}$ to $D$ extends 
the restriction of 
$A^{1/2}$ to $D(A)$. To prove this let $\xi$ be some 
element of $D(A)$. Since $D$ is a core for $A$ there 
is a sequence $\xi_0, \xi_1 \dots$ of elements of 
$D$ converging to $\xi$ and at the same time such that
$A\xi_0, A\xi_1 \dots$ converges to $A\xi$. Since 
$A^{1/2}$ has a bounded inverse it follows from this 
that $A^{1/2}\xi_0, A^{1/2}\xi_1 \dots$ vonverges 
to $A^{1/2}\xi$. Since $\xi$ can be chosen otherwise arbitrarily
it follows that the closure of
the restriction of $A^{1/2}$ to $D$ extends 
the restriction of 
$A^{1/2}$ to $D(A)$ and hence that $D$ is a core for $A^{1/2}$.
Hence for any $\xi \in D(A^{1/2})$ there is a 
a sequence $\xi_0, \xi_1 \dots$ in $D$ converging 
to $\xi$ and at the same time such that 
$A^{1/2}\xi_0, A^{1/2}\xi_1 \dots$ is converging to 
$A^{1/2}\xi$. Hence by (\ref{perturbB0}) along with the 
completeness of $X$ follows the convergence of the sequence $B_0\xi_1,
B_0\xi_2 \dots$ to some element $\bar{B}\xi$ of $X$ and
\begin{equation} 
\| 
\bar{B} \xi \|^2 
\leqslant a_0^2 \, \| A^{1/2} \xi \|^2 + b_0^2 \, \| \xi \|^2 \, \, .
\end{equation}
Moreover 
if  $\xi^{\, \prime}_0, \xi^{\, \prime}_1 \dots$ is another sequence 
having the same properties as $\xi_0, \xi_1 \dots$
by (\ref{perturbB0}) follows that
\begin{equation}
\bar{B}\xi = 
\lim_{n \rightarrow \infty}
B_0 \xi_n =
\lim_{n \rightarrow \infty}
B_0 \xi^{\, \prime}_n \, \, .
\end{equation} 
From this it easily seen that by defining 
\begin{equation}
\bar{B} := (D(A^{1/2}) \rightarrow X \, , \, 
\xi \mapsto \bar{B} \xi)
\end{equation}
there is also given a linear map. Hence the existence of 
a linear extension of $B_0$ satisfying (\ref{perturbbarB0}) 
is shown. Moreover from the definition it is obvious 
that $\bar{B}$ is symmetric if $B_0$ is in addition symmetric. 
If on the other hand $\bar{B}_0$ is a linear extension 
of $B_0$ satisfying (\ref{perturbbarB0})
and $\xi$ and $\xi_1 , \xi_2$ are as above 
from (\ref{perturbbarB0}) follows that 
\begin{equation}
\bar{B}_0 \xi = \lim_{n \rightarrow \infty} B_0 \xi_n \, \, .
\end{equation}
Finally, since $\xi$ can be chosen otherwise arbitrarily 
from this follows $\hat{B}_0 = \hat{B}$.$_{\square}$
\section{Discussion and results}
This paper provides a rigorous framework for 
the description of linearized adiabatic lagrangian perturbations 
and stability of differentially rotating newtonian stars
using semigroup theory. Problems of a 
previous framework by Dyson and Schutz 
are overcome and a basis for a rigorous analysis of the stability of 
such stars is provided. The spectrum of the oscillations is shown to 
coincide with the spectrum of an operator polynomial whose
coefficients can be read off from the equation governing the 
oscillations about the equilibrium 
configuration. Moreover, for the first time sufficient criteria 
for stability are given in form of inequalities for the coefficients
of that polynomial. These show that a negative canonical energy of 
the star does not necessarily indicate instability.
\newline

It is still unclear whether these criteria are strong enough to
prove stability for realistic stars. On the other hand
the second criterium has been sucessfully applied in the 
(on first sight seemingly unrelated case of the) stability 
discussion of the Kerr metric where the master equation governing 
perturbations is of the form (\ref{basiceq}), too.\cite{beyer4}
Another similarity of that case to the cases considered here
is the fact that the corresponding operators 
$C^{\, \prime}$ and $B^{\, \prime \, 2}$ there are such that 
$C^{\, \prime} - (1/4)B^{\, \prime \, 2}$ is positive whereas
here this combination is semibounded as has been shown 
in DS. 
\newline

Also the  
determination of the spectrum of the operator polynomial
$C^{\, \prime} - \lambda B^{\, \prime} + \lambda^2,
\lambda \in {\mathbb{C}}$ for some special case would be 
very useful. It is 
likely that this cannot be done for a physically relevant 
case. But it is also likely
that the outcome to {\it qualitative} questions like
\begin{itemize} 
\item
Does one have uniform stability in $m$? 
\item
Does a continuous part occur in the oscillation spectrum? 
\end{itemize}
only depends on {\it structural} properties of the operators 
$C^{\, \prime}$  and $B^{\, \prime}$.
So from $C^{\, \prime}$ probably only 
the highest order derivatives are relevant and  
details of the equation of state should be unimportant. 
From this point of view even the highly idealized case 
of a spherical background model with a truncated
$C^{\, \prime}$ along with a non constant 
velocity field $v$ would be interesting to 
consider.

\end{document}